# The Early Universe and Planck's Radiation Law

Rainer Collier

Theoretisch-Physikalisches Institut, Friedrich-Schiller-Universität Jena

Max-Wien-Platz 1, 07743 Jena, Germany

E-mail: rainer@dr-collier.de

**Abstract:** The classical Friedmann-Lemaître equations are solved using a corrected version of Planck's radiation law. The function curves of the scale parameter $a(t)$ and the variations with temperature $a(T)$ and $t(T)$ are given. It is shown that a reasonable cosmological evolution is only possible in case of flat spatial slices $(k=0)$. The initial singularity is avoided. Horizon and flatness problems do not exist. For temperatures $k_B T \ll E_*$ ($E_*$ Planck energy), the equations yield the usual course of expansion of the standard FLRW model for a radiation universe with $k = 0$ and $p = u(T)/3$.

## 1 Introduction

Results of astronomical observations made in recent years prove that the observable part of the universe is, over sufficiently large distances, almost homogeneously and isotropically filled with matter and radiation and has undergone a dynamic structural development. The redshift of electromagnetic radiation emitted by distant galaxies, the structure of the cosmic (photon) microwave background (CMB), the frequency of the element $He^4$ and other discoveries of the recent years [1] are well in agreement with the presently accepted standard model of a universe [25], [26], [27] that is expanding in space and time. Therefore, the main access to a description of the dynamics of the universe is Einstein's theory of gravitation for homogeneous and isotropic space-times with a suitable equation of state for the matter contained in it. It is this Friedmann-Lemaître-Robertson-Walker (FLRW) model that yields the fundamental equations of cosmic development, here called Friedmann equations for brevity. They have extended our understanding of the past and the future of the universe and ,of how its large-scale structure have come about. Problems related to the very early start of cosmic development (the big bang singularity), such as the horizon, flatness and monopole problems etc., are overcome, as a rule, with the aid of the inflation concept.

In this report, a corrected Planck's radiation law [2] as an equation of state for the photon gas is substituted in the Friedmann equations. For extremely high temperatures $k_B T \gg E_*$ ($E_*$ Planck energy), cosmic expansion starts, without any singular behaviour, from a Planck region of the size $L_*$ (Planck length). There is neither a horizon nor a flatness problem. Only a universe make sense having a curvature parameter $k = 0$. For lower temperatures $k_B T \ll E_*$, all curve shapes pass over into those of the well-known standard FLRW model with Planck's photon equation of state $p = u(T)/3$.

## 2 The fundamental equations of cosmology

A homogeneous and isotropic universe has a four-dimensional geometry, which is described by the Robertson-Walker line element

$$ds^2 = g_{ik} dx^i dx^k = c^2 dt^2 - a^2 \left( \frac{dr^2}{1-kr^2} + r^2 d\Omega^2 \right) \tag{2.1}$$

in which the scale parameter $a(t)$ is a function of the cosmic time $t$ only. The energy momentum tensor $T_{ik}$ must have a form adapted to the cosmological situation (2.1):

$$T_{ik} = \begin{pmatrix} e \cdot g_{00} & 0 \\ 0 & -p \cdot g_{\alpha\beta} \end{pmatrix}, \tag{2.2}$$

with $g_{00}$ and $g_{\alpha\beta}$ to be read from (2.1). In the tensor (2.2), $p(t)$ is the pressure function and $e(t) = \mu(t) + u(t)$ is the rest energy density, which is composed of the energy density of the rest masses $\mu(t)$ and the internal energy density $u(t)$. We realize that this ansatz does not require any additional assumptions, such as restriction to an ideal fluid. For example, $T_{ik}$ may also include matter viscosities [3] or be the energy momentum tensor of a scalar field $\Phi(t)$ with $p < 0$ [4].

In the metric (2.1) with energy momentum tensor (2.2), the Einstein equations then lead to the Friedmann equations

$$\left(\frac{\dot{a}}{a}\right)^2 + \frac{k \cdot c^2}{a^2} = \frac{1}{3}\left(\frac{8\pi G}{c^2} e + \Lambda \cdot c^2\right) \quad \rightarrow \quad (0,0) - Equation, \tag{2.3}$$

$$\frac{\ddot{a}}{a} + \left(\frac{\dot{a}}{a}\right)^2 + \frac{k \cdot c^2}{a^2} = -\frac{8\pi G}{c^2} p + \Lambda \cdot c^2 \quad \rightarrow \quad (\alpha,\alpha) - Equations. \tag{2.4}$$

Here, $k = (0, \pm 1)$ is the curvature parameter, $c$ the velocity of light in vacuum, G Newton's gravitational constant, $\Lambda$ the cosmological constant and $\dot{a} = da/dt$. In addition, we consider the integrability condition $T^{ik}{}_{;k} = 0$. In accordance with the above ansatz, the $(\alpha)$-equations yield only $\partial p/\partial x^\alpha = 0$, whereas the $(0)$-equation constitutes the equation of motion of the cosmic matter:

$$\dot{e} + 3\left(\frac{\dot{a}}{a}\right)(e+p) = 0 \quad . \tag{2.5}$$

Equation (2.4) results from differentiation of (2.3) and the use of (2.5). It is thus sufficient to solve the Friedmann equation (2.3) in compliance with the integrability condition (2.5). By inserting (2.3) into (2.4), we additionally obtain a statement about the acceleration behaviour of the universe :

$$\ddot{a} = -\frac{4\pi G}{3c^2}(e+3p) \cdot a \quad . \tag{2.6}$$

As long as the strong energy condition $(e+3p) > 0$ is satisfied, an existing expansion will decelerate, while otherwise it will accelerate.

For the following, let us absorb the cosmological constant $\Lambda$ on the right of (2.3) and (2.4) by the substitutions

$$e + \frac{\Lambda c^4}{8\pi G} \rightarrow e \quad , \quad p - \frac{\Lambda c^4}{8\pi G} \rightarrow p \quad . \tag{2.7}$$

This substitution leaves the equation of motion (2.5) invariant in form. Thus, all matter, including its dark $\Lambda$ – components, is included in $e(t)$ and $p(t)$ [26]. Therefore, the fundamental equations of cosmology we use are the following differential equations:

$$\left(\frac{\dot{a}}{a}\right)^2 + \frac{kc^2}{a^2} = \frac{8\pi G}{3c^2} \cdot e \qquad \textit{Friedmann equation} \quad , \tag{2.8}$$

$$\dot{e} + 3\left(\frac{\dot{a}}{a}\right)(e+p) = 0 \qquad \textit{matter equation of motion} \quad , \tag{2.9}$$

$$\frac{\ddot{a}}{a} = -\frac{4\pi G}{3c^2}(e+3p) \qquad \textit{acceleration equation} \quad . \tag{2.10}$$

Once we have chosen an equation of state $p = p(e)$, we determine the wanted functions $a(t)$ and $e(t)$ from the two equations (2.8) and (2.9), which enables us to discuss also, from (2.10), the acceleration behaviour of the universe.

## 3 The equation of state $p = w \cdot e$

There is one interesting conclusion we can immediately draw from the fundamental equations (2.8) and (2.9). Should a constant energy density $e = const$ occur in any period of time, however short, during cosmic expansion, the equation of state $p = -e$ follows from (2.9), and thus, $\ddot{a} > 0$ from (2.10). In this phase of accelerated expansion, the linear extension of a $(k = 0)$ universe according to (2.8) increases exponentially (cf. (3.2)). This is the scenario of inflation!

Let us now regard the well-known solutions for equations of state of the form

$$p = w \cdot e \ , \qquad w = const \ . \qquad (3.1)$$

The corresponding solution of the Friedmann equation (2.8) for $k = 0$ is known to be

$$a = \begin{cases} const \cdot t^{\frac{2}{3(1+w)}} \ , & w \neq -1 \ , \\ const \cdot \exp(H \cdot t) \ , & w = -1 \ , \end{cases} \qquad (3.2)$$

where $H = \sqrt{(8\pi G/3c^2)e}$. The behaviour of the energy density $e$ for all values $w$ results from (2.9),

$$e(a) = const / a^{3(1+w)} \qquad (3.3)$$

and, with (3.2) inserted in (3.3), so does its time dependence

$$e(t) = \begin{cases} const / t^2 \ , & w \neq -1 \ , \\ const \ , & w = -1 \ . \end{cases} \qquad (3.4)$$

It can be seen that the time behaviour of the energy density $e$ is independent of the constant $w$ value, and so is, therefore, the Hubble parameter $H(t)$,

$$H(t) = \left(\frac{\dot{a}}{a}\right) = \sqrt{\frac{8\pi G}{3c^2} e(t)} = \begin{cases} const / t \ , & w \neq -1 \ , \\ const \ , & w = -1 \ . \end{cases} \qquad (3.5)$$

Here we may note that the solutions $a(t)$ for the curvature cases $k = \pm 1$ in passage to the limit $t \to 0$ also show the behaviour (3.2) of the solution for $k = 0$.

**4 Problems related to the equation of state $p = w \cdot e$**

The Friedmann model together with the classical equation of state $p = w \cdot e$ and the solution (3.2) with $w > -(1/3)$ leaves some questions open in passage to the limit $t \to 0$, irrespective of the curvature parameter $k$, (see also [25]). Below, the most important problems of the standard cosmological model are briefly outlined again.

4.1 *The big bang problem*

From the acceleration equation (2.10) there follows $\ddot{a} < 0$, as long as the strong energy condition $e + 3p > 0$ is satisfied. For cosmic models with $p = w \cdot e$, this applies in case of constant $w > -(1/3)$. Then, for an expanding flat universe, the curve $a(t)$ is concave down and rising monotonously. Hence, at some time in the past, $a$ must have been equal to zero. And because, according to (3.2), for $w > -(1/3)$, the velocity $\dot{a} = const \cdot t^{-\delta}$ with $\delta = [(1+3w)/3(1+w)] > 0$ diverges at $t \to 0$, the cosmic development starts with a big bang,

$$a(t) \to 0 \quad , \quad \dot{a}(t) \to \infty \quad , \quad \text{for } t \to 0 \ . \tag{4.1}$$

Also, according to (3.4), the energy density $e(a)$ approximates infinity.

*4.2 The horizon problem*

Let us restrict our investigation to the case of gravitational attraction $e + 3p > 0$, i.e. to $w > -(1/3)$. In this case, $a(t)$ as a function of time is given by (3.2). The size of the visible part of the universe, the particle horizon $D$, is given by the length of the light path from the big bang $t = 0$ up to time $t$ [5],

$$D(t) = a(t) \int_0^t \frac{c}{a(\tau)} d\tau \ . \tag{4.2}$$

If $a(t)$ is described by a power law according to (3.2), the size $D$ of the horizon region is calculated to be

$$D(t) = \frac{3(1+w)}{1+3w}(ct). \tag{4.3}$$

For $w = 1/3$, the well-known terms $a(t) = const \cdot \sqrt{t}$ and $D(t) = 2ct$ appear. To guarantee causality, the linear extension $a$ of the universe should always be smaller than the particle horizon $D$:

$$const \cdot t^{\frac{2}{3(1+w)}} < t \ . \tag{4.4}$$

Since, for $w > -(1/3)$, always $2/3(1+w) < 1$, there is, for $t \to 0$, a moment $t$ from that one the inequality (4.4) is violated, and therefore the extension $a(t)$ of the universe exceeds the linear measure $D(t)$ of the horizon. This raises the question how the homogeneity and isotropy observed today in the cosmic microwave background could have come about, if in an early phase of cosmic expansion certain regions of the universe were not causally connected.

*4.3 The flatness problem*

Let us transform the Friedmann equation (2.8) by introducing the Hubble parameter $H(t)$ and the density parameter $\Omega(t)$,

$$H = \frac{\dot{a}}{a} \quad , \quad \Omega = \frac{\rho}{\rho_c} \quad , \quad \rho = \frac{e}{c^2} \quad , \quad \rho_c = \frac{3}{8\pi G} H^2 \ . \tag{4.5}$$

where $\rho_c$ denotes the critical density, which in a flat universe with $k=0$ is calculated from (2.8). With these definitions, the Friedmann equation can also be written as

$$|\Omega - 1| = |k|\frac{c^2}{a^2 H^2} = |k|\frac{c^2}{\dot{a}^2} \quad . \tag{4.6}$$

Since for $t \to 0$, the solutions $a(t)$ of cosmic models with $k = \pm 1$ behave like the solution for $k=0$ from (3.2), we insert this $\dot{a}$ into (4.6) and get

$$|\Omega - 1| = |k| const \cdot t^{2\delta} \tag{4.7}$$

with $\delta > 0$ from (4.1). For $w = 1/3$, there follows the well-known value $|\Omega - 1| = |k| \cdot const \cdot t$. For $w > -(1/3)$, the right-hand term of (4.7) moves with increasing time $t > 0$ from $\Omega = 1$ towards rising values. However, our universe of today still has approximately the value $\Omega \simeq 1$. Obviously, the cosmic model with $w > -(1/3)$ is unsuitable for the early phase of cosmic development, or it exactly satisfies $k = 0$ for all times, which needs to be well founded on physics.

## 5 The new radiation universe

A corrected Planck's radiation law derived in an earlier report [2] will now be applied to the early phase of cosmic development. Let us first repeat the main lines of thought that led to the new equations of state for the photon gas, $p = p(T)$ and $u = u(T)$. The starting point was the limitation of the size of the energy quanta $\hbar \omega$, originating from a very simple quantizing model of the Friedmann universe containing dust and radiation. As a result of these investigations, the following universally valid hypothesis was proposed:

*As a consequence of the discrete micro-structure of space-time manifold, quantum jumps $\Delta E$ between the energy eigenstates of a quantum system are universally limited to*

$$\Delta E \leq E_* \quad , \tag{5.1}$$

with $E_* = M_* c^2 = \alpha M_{Pl} c^2$ and $M_{Pl} = \sqrt{\hbar c / G}$.

In other words, it is assumed that the size of energy quanta in nature has a universal limit given by the Planck energy $E_*$. The quantity $E_*$ or $M_*$, respectively, has to be determined either experimentally in the lab or from astrophysical observations. In (5.1) we first use $\alpha = 1$, so that $M_* = M_{Pl}$. Therefore, for the gap between the equidistant energy levels of the harmonic oscillator (and, thus, for the size of the energy portions of the electromagnetic radiation having the frequency $\omega$), it applies a universal limit of

$$\hbar \omega \leq E_* \quad . \tag{5.2}$$

*5.1 General thermodynamics of the photon gas*

Before we apply the hypothesis (5.1) to the photon gas, let us write down the essential thermodynamic relations. It is known that, for a photon gas, the chemical potential $\mu$ vanishes, and the densities of the extensive thermodynamic potentials are functions of temperature only,

$$\mu = 0 \quad , \quad F = f(T) \cdot V \quad , \quad U = u(T) \cdot V \quad , \quad S = s(T) \cdot V \quad . \tag{5.3}$$

In the equilibrium thermodynamics to be assumed here, the differential of the free energy $F$ has the form

$$dF = -S dT - p dV \quad , \quad S = -\left(\frac{\partial F}{\partial T}\right)_V \quad , \quad p = -\left(\frac{\partial F}{\partial V}\right)_T \quad . \tag{5.4}$$

Introducing the abbreviation $f' = df/dT$ etc. and considering $F = U - TS$, we get

$$s = -f' \quad , \quad p = -f = -(u - Ts) \quad , \quad p + u = Ts \quad . \tag{5.5}$$

Because of $Ts = -Tf' = Tp'$, the fundamental relation is

$$T p' = p + u \quad , \tag{5.6}$$

which establishes a relationship between the thermal and the caloric equation of state of the photon gas. This generalized equation of state must be satisfied by any photon gas.

*5.2 Thermodynamic potentials of the new radiation law*

If we now apply the hypothesis (5.1) to the (thermalized) photon gas, the relevant thermodynamic potentials – viz. the free energy $F$, the internal energy $U$, the number of particles $N$, the entropy $S$ and the pressure $p$ – take the following form [2]:

$$F = \frac{g_s \cdot V}{2\pi^2 (\hbar c)^3}(k_B T) \int_0^{E_*} \frac{(\hbar\omega)^2}{\left(1 - \frac{\hbar\omega}{E_*}\right)} \ln\left[1 - \left(1 - \frac{\hbar\omega}{E_*}\right)\exp\left(-\frac{\hbar\omega}{k_B T}\right)\right] d(\hbar\omega) \tag{5.7}$$

From $F = -k_B T \ln Z$ we read the integral of state $Z$ and compute the internal energy from $U = -(\partial/\partial\beta)\ln Z$ with $\beta = 1/k_B T$, which yields

$$U = \frac{g_s \cdot V}{2\pi^2 (\hbar c)^3} \int_0^{E_*} \frac{(\hbar\omega)^3}{\exp\left(\frac{\hbar\omega}{k_B T}\right) - \left(1 - \frac{\hbar\omega}{E_*}\right)} d(\hbar\omega) \quad , \tag{5.8}$$

$$N = \frac{g_s \cdot V}{2\pi^2 (\hbar c)^3} \int_0^{E_*} \frac{(\hbar\omega)^2}{\exp\left(\frac{\hbar\omega}{k_B T}\right) - \left(1 - \frac{\hbar\omega}{E_*}\right)} d(\hbar\omega) , \qquad (5.9)$$

$$S = \frac{1}{T}(U - F) \quad , \quad p = -\frac{F}{V} , \qquad (5.10)$$

wherein $T$ denotes the temperature, $V$ the volume, $g_s$ the spin degeneration factor, $2\pi\hbar = h$ Planck's constant and $k_B$ Boltzmann's constant. The quantity essential for the time history $a(t)$ of the scale factor in the Friedmann equation (2.8) is the energy density $e$, which for the photon gas consists only of internal energy $e = u = U/V$:

$$u(T) = \frac{g_s}{2\pi^2 (\hbar c)^3} \int_0^{E_*} \frac{(\hbar\omega)^3}{\exp\left(\frac{\hbar\omega}{k_B T}\right) - \left(1 - \frac{\hbar\omega}{E_*}\right)} d(\hbar\omega) . \qquad (5.11)$$

Introducing the spectral energy density $\bar{u}$ and the mean energy $\bar{\varepsilon}$ at the level $\hbar\omega$ and at the temperature $T$, we find from (5.11) that

$$u(T) = \int_0^{E_*/\hbar} \bar{u}(\omega, T) \, d\omega \quad , \quad \bar{u} = \frac{g_s \omega^2}{2\pi^2 c^3} \bar{\varepsilon} \quad , \quad \bar{\varepsilon}(\omega, T) = \frac{\hbar\omega}{\exp\left(\frac{\hbar\omega}{k_B T}\right) - \left(1 - \frac{\hbar\omega}{E_*}\right)} . \qquad (5.12)$$

One verifies that the new equations of state (5.10) and (5.11) for $p$ and $u$ satisfy the important differential equation of state (5.6). It can also be seen that Planck's radiation formulae appear for $E_* \to \infty$.

*5.3 Properties of energy density and pressure*

Before we look for solutions of the Friedmann equations with the new radiation equations of state (5.7) to (5.10), let us state some properties of the new radiation laws. In the earlier study [2] we investigated the behaviour of all relevant thermodynamic functions and drew up diagrams of their curves. Here we restrict ourselves to the properties of energy density $u$ and pressure $p$. For this purpose we introduce the dimensionless variables $z$ and $\theta$,

$$z = \frac{\hbar\omega}{k_B T} \quad , \quad \theta = \frac{k_B T}{E_*} \quad , \quad \theta \cdot z = \frac{\hbar\omega}{E_*} \qquad (5.13)$$

and use the following Planck quantities:

$$E_* = M_* c^2 = \sqrt{\frac{\hbar c}{G}} \cdot c^2 \quad , \quad u_* = \frac{E_*}{V_*} \quad , \quad V_* = 2\pi^2 L_*^3 \quad , \quad L_* = \sqrt{\frac{\hbar G}{c^3}} = t_* \cdot c . \qquad (5.14)$$

The energy density $u$ from (5.11) and the pressure $p$ from (5.10) with (5.7) can then be written as

$$u(\theta) = g_s u_* \theta^4 \int_0^{1/\theta} \frac{z^3 \, dz}{\exp(z) - (1 - \theta z)} \quad , \tag{5.15}$$

$$p(\theta) = -g_s u_* \theta^4 \int_0^{1/\theta} \frac{z^2}{(1 - \theta z)} \ln\left[1 - (1 - \theta z)\exp(z)\right] dz \quad . \tag{5.16}$$

It is easy to see that Planck's radiation formulae appear for $\theta \to 0$. A form of (5.15), (5.16) that is well suitable for numerical analyses can be obtained by the substitution of $x = \theta \cdot z$:

$$u(\theta) = g_s u_* \int_0^1 \frac{x^3 \, dx}{\exp\left(\frac{x}{\theta}\right) - (1 - x)} \quad , \tag{5.17}$$

$$p(\theta) = -g_s u_* \theta \int_0^1 \frac{x^2}{(1 - x)} \ln\left[1 - (1 - x)\exp\left(\frac{x}{\theta}\right)\right] dx \quad . \tag{5.18}$$

We realize that the ratio $w(\theta) = p(\theta)/u(\theta)$ now varies with temperature (it will rise with increasing temperature). Only for $k_B T \ll E_*$, i.e. $\theta \ll 1$, it will reach the Planck limit $\lim_{\theta \to 0} w(\theta) = 1/3$. The approximation formulae established in [2] for $u, p$ and $w$ have the form ($\sigma$ Stefan-Boltzmann constant, $\varsigma(x)$ Riemann zeta function),

$\theta \ll 1 \, (k_B T \ll E_*)$:

$$u(\theta) = u_{Pl}(1 - \eta_{(4)}\theta) \quad , \quad u_{Pl} = g_s u_* \frac{\pi^4}{15} \theta^4 = \frac{4}{c}\sigma T^4 \quad , \tag{5.19}$$

$$\eta_{(4)} = 4(1 - \frac{\varsigma(5)}{\varsigma(4)}) \approx 0.17 \quad , \quad g_s = 2 \quad , \quad \sigma = \frac{2\pi^5 k_B^4}{15 h^3 c^2} \quad ,$$

$$p(\theta) = p_{Pl}(1 + \frac{3}{4}\eta_{(4)}\theta) \quad , \quad p_{Pl} = \frac{1}{3} u_{Pl} \quad , \tag{5.20}$$

$$w(\theta) = \frac{p(\theta)}{u(\theta)} = \frac{1}{3}\left(1 + \frac{1}{4}\eta_{(4)}\theta\right) \quad . \tag{5.21}$$

$\theta \gg 1 \, (k_B T \gg E_*)$:

$$u(\theta) = \frac{g_s}{3} u_* \left(1 - \frac{1}{\theta}\right) \quad , \quad g_s = 2 \quad , \tag{5.22}$$

$$p(\theta) = g_s u_* \left( C_\infty \cdot \theta - \frac{1}{3} \right) \quad , \quad C_\infty = \frac{\pi^2}{6} - \frac{5}{4} \approx 0.395 \quad , \tag{5.23}$$

$$w(\theta) = \frac{p(\theta)}{u(\theta)} = 3C_\infty \cdot \theta + (3C_\infty - 1) \quad . \tag{5.24}$$

For $\theta \to \infty$, the energy density $u(\theta)$ has the limit $u_\infty = g_s u_*/3 = (2/3)u_*$ (figure 1). This limit behaviour for $u(\theta)$ at high temperatures was to be expected, as the new radiation laws (5.7) to (5.10) are based on the hypothesis (5.1). Even with the highest possible thermal excitation ($\theta \to \infty$) of the photon gas, the mean energy $\bar{\varepsilon}$ in (5.12) will not exceed the energy limit $E_*$. Hence, however, the energy density $u(\theta)$ must always remain finite as well. $\lim_{\theta \to \infty} u(\theta) = u_\infty = const$ is those property that makes the new radiation law (5.15) or (5.17) interesting for application to the early phase of the universe. As the start of cosmic evolution is characterized by extremely high temperatures $k_B T \gg E_*$, the energy density $u$ at that phase will take a near-constant course. According to the Friedmann equation (2.8) with $k = 0$, an energy density that is constant over time causes exponential growth of the scale factor $a(t)$. As, on the other hand, decreasing temperatures $k_B T \ll E_*$ obey Planck's radiation laws again, the scale factor $a(t)$ for this temperature range will pass over in the usual time behaviour of the standard model.

With increasing temperature $\theta \to \infty$, the pressure $p(\theta)$ will asymptotically approach the straight line (5.23), whose intersection with the $p$ axis is at $p = -u_\infty$ (figure 2). Hence, with increasing temperature, the ratio $w(\theta) = p(\theta)/u(\theta)$ asymptotically rises towards this straight line (figure 3). Studies for very low temperatures $k_B T \ll E_*$ leading to $w = w(T)$, have already been conducted in [6], [7] and [8].

It should further be noted that other authors have also reported about limit energy densities $u_{crit} \sim u_*$ occurring in a corrected Friedmann equation of the form $H^2 = u(1 - u/u_{crit})$. These limits result from different ansatzes for extending the fundamental physics (string theory, LQG, DSR, GUP, deformed dispersion relations, … ). A selection out of it can be found in [9] to [23]. Radiation laws with modified equations of state have been treated in [28], [29,], [30].

*5.4 Approximation solutions of the Friedmann equation*

Let us now find solutions of the differential equations (2.8) and (2.9) for the early universe, using the new photon equations of state (5.15) and (5.16). Let these be considered, at the same time, as equations of state for an extremely relativistic particle gas, the particle energies therein are large compared to the rest mass energies. As astronomical observations with $\Omega \approx 1$ indicate a spatially flat universe, we first look for solutions with $k = 0$.

We introduce suitable normalizing factors,

$$\lim_{t \to 0} a(t) = a_0 \quad , \quad \lim_{t \to 0} \dot{a}(t) = c \quad , \quad \lim_{\theta \to \infty} u(\theta) = u_\infty \quad .$$

Their values can be computed from the Friedmann equations (2.8), (2.9) at the time $t=0$ ($\theta = \infty$), the constant $u_\infty$ follows for $\theta \to \infty$ ($t=0$) from (5.22),

$$a_0 = \sqrt{\frac{3c^4}{8\pi G \cdot u_\infty}} = \sqrt{\frac{9\pi}{8}} L_* = 1.880\, L_* \quad , \quad t_0 = \frac{a_0}{c} = \sqrt{\frac{9\pi}{8}} t_* = 1.880\, t_* \quad , \quad u_\infty = \frac{2}{3} u_* \quad . \quad (5.25)$$

In the following, all functions $a, \dot a, u, p$ etc. are related to these normalizing constants, i.e., we calculate in a system of units $[a_0, t_0, u_\infty]$. After that we set

$$a_0 = 1 \quad , \quad t_0 = 1 \quad , \quad u_\infty = 1 \quad . \tag{5.26}$$

The Friedmann equations for the flat universe read now

$$\left(\frac{\dot a}{a}\right)^2 = u \quad , \quad \dot u + 3\left(\frac{\dot a}{a}\right)(u+p) = 0 \quad , \tag{5.27}$$

and the equations of state $u(\theta), p(\theta)$ as obtained from (5.15), (5.16) have the form

$$u(\theta) = 3\,\theta^4 \int_0^{1/\theta} \frac{z^3}{\exp(z)-(1-\theta z)} dz \quad , \quad p(\theta) = -3\,\theta^4 \int_0^{1/\theta} \frac{z^2}{1-\theta z} \ln[1-(1-\theta z)\exp(z)] \, dz \quad . \tag{5.28}$$

As pressure and energy density are given as functions of temperature $u = u(\theta)$, $p = p(\theta)$, we rewrite the equations (5.27) using the temperature parameter $\theta = k_B T / E_*$,

$$\dot a = \frac{da}{d\theta}\bigg/\frac{dt}{d\theta} = \frac{a'}{t'} \quad , \quad \dot u = \frac{u'}{t'} \quad . \tag{5.29}$$

Now depending on the temperature parameter $\theta$, the Friedmann equations are

$$t' = \left(\frac{a'}{a}\right)\frac{1}{\sqrt{u}} \quad , \quad u' + 3\left(\frac{a'}{a}\right)(u+p) = 0 \quad , \tag{5.30}$$

from which immediately the following solutions can be derived,

$$t(\theta) = -\frac{1}{3}\int_\infty^\theta \left(\frac{u'}{u+p}\right)\frac{1}{\sqrt{u}}\, d\vartheta \quad , \quad a(\theta) = \exp\left[-\frac{1}{3}\int_\infty^\theta \left(\frac{u'}{u+p}\right) d\vartheta\right] \quad . \tag{5.31}$$

Let us first inspect these solutions in their thermal boundary regions.

$\theta \ll 1$ ($k_B T \ll E_*$):

In the case of sufficiently low temperatures compared with the Planck temperature $k_B T = E_*$, the equations of state of pressure and energy density (5.15) and (5.16) turn into those of Planck's photon gas. Considering our system of units (5.26), Planck's radiation law (5.19) reads as

$$u = \frac{\theta^4}{(2A)^2} \quad , \quad A = \frac{\sqrt{5}}{2\pi^2} \quad , \quad p = \frac{1}{3}u \quad . \tag{5.32}$$

Substituting (5.32) into the solution structure (5.30) results in

$$t' = \left(-\frac{1}{\theta}\right)\left(\frac{2A}{\theta^2}\right) = -2\frac{A}{\theta^3} \quad , \quad t = \frac{A}{\theta^2} \quad , \quad (\ln a)' = -\frac{1}{\theta} \quad , \quad a = \frac{B}{\theta} \quad . \tag{5.33}$$

The solutions $t \cdot \theta^2 = A$ and $a \cdot \theta = B$ are the solutions of the $(k=0)$ standard model for Planck's photon gas. The constant $A$ is defined in (5.32), whereas the constant $B$ is yet indefinite. From (5.32) and (5.33), there follows the time behaviour already known from (3.1) to (3.3) for $w = 1/3$:

$$a(t) = \frac{B}{\sqrt{A}} \cdot \sqrt{t} \quad , \quad u(t) = \frac{1}{4} \cdot \frac{1}{t^2} \quad , \quad \theta(t) = \sqrt{A} \cdot \frac{1}{\sqrt{t}} \quad . \tag{5.34}$$

$\theta \gg 1 \, (k_B T \gg E_*)$ :

In the case of extremely high temperatures compared with the Planck temperature $k_B T = E_*$, the pressure and energy equations of state have the form (5.22), (5.23). With the system of units (5.26), these equations turn into

$$u = 1 - \frac{1}{\theta} \quad , \quad p = 3C_\infty \theta - 1 \quad . \tag{5.35}$$

Using the solutions (5.31) yields the approximation

$$t' = \left(-\frac{1}{9C_\infty}\right) \cdot \frac{1}{\theta^3}, \quad t = \frac{1}{18C_\infty} \cdot \frac{1}{\theta^2}, \quad (\ln a)' = \left(-\frac{1}{9C_\infty}\right) \cdot \frac{1}{\theta^3}, \quad a = \exp\left(\frac{1}{18C_\infty} \cdot \frac{1}{\theta^2}\right). \tag{5.36}$$

The solution $t \cdot \theta^2 = 1/18 C_\infty = const$ has the same behaviour as (5.33) in the low-temperature case, except for a different constant.

The scale parameter $a(t)$ shows the expected exponential behaviour for $t \to 0$, as can be seen in (5.36) in the transition to the time dependences,

$$a(t) = \exp(t) \quad , \quad u(t) = \exp\left(-\sqrt{18C_\infty} \cdot \sqrt{t}\right) \quad , \quad \theta(t) = \frac{1}{\sqrt{18C_\infty}} \cdot \frac{1}{\sqrt{t}} \quad . \tag{5.37}$$

*5.5 Exact solutions of the Friedmann equation*

We will show that the second equation in (5.30) can be solved exactly, no matter which pair of photon equations of state $u(\theta), p(\theta)$ is used. For this purpose we use the thermodynamic relation (5.6) in the system of units (5.26) and differentiate by $\theta$ :

$$\theta p' = u + p \quad , \quad \theta p'' = u' \quad . \tag{5.38}$$

Substitution in (5.30) yields

$$\left(\frac{a'}{a}\right) = -\frac{1}{3}\frac{u'}{u+p} = -\frac{1}{3}\frac{\theta p''}{\theta p'} = -\frac{1}{3}(\ln p')' \quad , \tag{5.39}$$

$$(\ln a^3)' + (\ln p')' = 0 \quad , \quad (a^3 \cdot p') = C = const. \tag{5.40}$$

Finally we can define the constant $C$ by transition to the limit $\theta \to \infty$ ($t \to 0$). With

$$\lim_{\theta \to \infty} a(\theta) = 1 \quad , \quad \lim_{\theta \to \infty} p'(\theta) = 3C_\infty \tag{5.41}$$

we get $C = 3C_\infty$. Hence, the exact solution of the second differential equation (5.30) has two versions (irrespective of the cosmological model $k = (0, \pm 1)$). With (5.38) and (5.40) arises

$$a^3 \cdot p' = 3 C_\infty \quad , \quad a^3 = \frac{3 C_\infty \theta}{u+p} \quad . \tag{5.42}$$

What is the physics behind this solution? If we look at the photon thermodynamics in chapter 5.1, we can see that, in the equilibrium case with $V \sim a^3$, the equations (5.42) simply means entropy conservation. The cosmological expansion is then, for some time, a strictly adiabatic process,

$$S = V \cdot s = const \cdot a^3 \cdot \frac{dp}{dT} = const \cdot (a^3 p') = const \quad . \tag{5.43}$$

The solution of the first equation (5.30) yields the time dependence of cosmic expansion. With (5.39), it is just a simple integration of known functions now,

$$t(\theta) = \int_\infty^\theta \left(\frac{a'}{a}\right) \frac{1}{\sqrt{u(\vartheta)}} d\theta = -\frac{1}{3}\int_\infty^\theta \left(\frac{p''}{p'}\right) \frac{1}{\sqrt{u(\vartheta)}} d\theta \quad . \tag{5.44}$$

The solution of structure (5.44) with (5.42), we can apply to any photon radiation formula satisfying the thermodynamic requirements (5.3) and (5.6). For Planck's radiation formulae (5.32), this is verified immediately. We get

$$u = \frac{\theta^4}{(2A)^2} \quad , \quad p = \frac{1}{3}u \quad , \quad p' = \frac{1}{3}\frac{\theta^3}{A^2} \quad , \quad p'' = \frac{\theta^2}{A^2} \tag{5.45}$$

and from this with (5.42) and (5.44) the solutions

$$a^3 = \frac{9C_\infty A^2}{\theta^3} \quad , \quad a \cdot \theta = (9C_\infty A^2)^{\frac{1}{3}} = B = const, \tag{5.46}$$

$$t = -\frac{1}{3}\int_\infty^\theta \left(\frac{3}{\vartheta}\right)\left(\frac{2A}{\vartheta^2}\right) d\vartheta = -2A\int_\infty^\theta \frac{d\vartheta}{\vartheta^3} = \frac{A}{\theta^2} \quad . \tag{5.47}$$

With constant B now defined in (5.46), we get with (5.32) to (5.34) then

$$a(t) = \frac{B}{\sqrt{A}} \cdot \sqrt{t} = 1,062 \cdot \sqrt{t} \quad , \quad A = \frac{\sqrt{5}}{2\pi^2} = 0.113 \quad , \quad B = \left(9 C_\infty A^2\right)^{\frac{1}{3}} = \left(\frac{45 C_\infty}{4\pi^4}\right)^{\frac{1}{3}} = 0.357 \, . \quad (5.48)$$

The solutions (5.42), (5.44), likewise, are applicable to the new radiation laws (5.28), which comprise Planck's laws for sufficiently low temperatures.

In the figures 4 - 12 the respective curves of the exact solutions for these new radiation laws are shown (red graphs), compared with the respective Planck curves (blue graphs). Shown are the variations with both temperature $a(\theta), t(\theta), u(\theta) \ldots$ and time $a(t), u(t), H(t) \ldots$ .

## 6 Initial behaviour, horizon and flatness problems

Let us examine some properties of the exact solutions of the Friedmann equations (5.31) with the new radiation laws (5.28) in the spatially flat universe ( $k = 0$ ).

As can be seen from the figures (4-12), only for extremely high temperatures $\theta = k_B T / E_* \to \infty$ (i.e. around the start phase of universe close to $t \simeq 0$) behave all relevant functions $a(\theta), t(\theta), \ldots$ distinctly different than the respective Planck functions. This is made clear by a short numerical table (also cf. figure 6).

| $\theta = k_B T / E_*$ | $t/t_0$ | $a_{Pl}/a_0$ | $a_{new}/a_0$ | $\frac{\Delta a}{a}$ [%] |
|---|---|---|---|---|
| $\infty$ | 0 | 0 | 1 | 100 |
| 1.0 | 0.10 | 0.34 | 1.08 | 68 |
| 0.1 | 10.35 | 3.41 | 3.60 | 5 |
| 0.03 | 124.99 | 11.87 | 11.93 | 0.5 |
| 0.01 | 1132.41 | 35.72 | 35.75 | 0.1 |
| 0.001 | $1.13 \cdot 10^5$ | 357.30 | 357.31 | 0.003 |

(6.1)

Here, $a_{new}(t)$ describes the variation of the scale factor $a(t)$ in the case of the new, corrected Planck's radiation law (5.28), and $a_{Pl}(t)$ the variation of $a(t)$ in the case of Planck's old radiation law in the version (5.32). The relative deviation

$$\frac{\Delta a}{a} = \frac{a_{new} - a_{Pl}}{a_{new}} \quad (6.2)$$

indicates the difference between the two curves in the lapse of time $t$. Already after a few 100 Planck times, the new $a_{neu}(t)$ curve approximates the old $a_{Pl}(t)$ curve up to a deviation of less than 1 %.

## 6.1 The initial behaviour

We examine the scale parameter $a(t)$ at $t = 0$. From the Friedmann equations (5.27) and their exact solution (5.42), together with the limit behaviour (5.35) for $u(\theta)$ and $p(\theta)$, the following initial behaviour results,

$$\lim_{t \to 0} a(t) = 1 \quad , \quad \lim_{t \to 0} \dot{a}(t) = 1 \quad . \tag{6.3}$$

In the common [m, kg, s] system, the values are

$$a(t = 0) = a_0 = \sqrt{\frac{9\pi}{8}} L_* = 1{,}880 \, L_* \quad , \quad \dot{a}(t = 0) = \frac{a_0}{t_0} = \frac{L_*}{t_*} = c \quad . \tag{6.4}$$

Therefore, with the new radiation laws based on the classical Einstein equations, there is a normal primordial start phase with finite initial conditions for the scale parameter $a(t)$. To find out whether there are singularities in the course of expansion nevertheless, we regard the acceleration equation of the Friedmann system of equations (2.10). Rewritten for the system of units in (5.26), it reads

$$\ddot{a} = -\frac{a}{2}(u + 3p) \tag{6.5}$$

Although for $\theta \to \infty$, $\lim_{\theta \to \infty} a = 1$ and $\lim_{\theta \to \infty} u = 1$, the equation (5.35) shows that $\lim_{\theta \to \infty} p \to \infty$. Hence, for $t \to 0$,

$$\lim_{t \to 0} \ddot{a} = -\infty \quad , \quad \lim_{t \to 0} R \sim \frac{6}{c^2}\left(\frac{\ddot{a}}{a}\right) + \cdots \to -\infty \quad . \tag{6.6}$$

The 4-dimensional Ricci scalar $R$ (despite $a(0) \neq 0$ and $\dot{a}(0) \neq 0$) has a genuine acceleration singularity, which is caused by the extremely high initial pressure $p \to \infty$ of the photon gas (see also [24]).

## 6.2 The horizon problem

It is known that at $t \to 0$ in Planck's photon universe, the particle horizon is shrinking faster than the linear dimension of the universe (cf. Fig. [8]). Therefore in the early universe all its regions cannot be in a causal relationship. However, this contradicts the largely homogeneous and isotropic structure of the cosmic microwave background as established by astrophysics.

The size of the particle horizon $D$ at the time $t$ was defined by the path length of a light signal emitted at the start of cosmic expansion at $t = 0$. If the universe starts its evolution in time at $t = 0$ with a scale parameter of finite size, $a(t = 0) = 1$, the evolution of the horizon should start there as well, $D(t = 0) = 1$ (in the units of (5.26)). In the time after, then,

$$D(t) = 1 + a(t) \int_0^t \frac{1}{a(\tau)} d\tau \geq a(t) \quad , \quad \frac{1}{a(t)} + \int_0^t \frac{1}{a(\tau)} d\tau \geq 1 \tag{6.7}$$

must always be valid. From the second inequality we get, by differentiation for $t$, the condition

$$-\frac{\dot{a}}{a^2} + \frac{1}{a} \geq 0 \quad , \quad -\frac{\dot{a}}{a} + 1 \geq 0 \quad , \quad \frac{\dot{a}}{a} \leq 1 \quad . \tag{6.8}$$

In the case of the Friedmann equation (5.27), $\dot{a}/a = \sqrt{u}$. In $0 \leq t < \infty$, the function $u(t)$ is a monotonously decreasing one, as with (5.27) we can see from

$$u(0) = 1, \quad u(t) \geq 0 \quad , \quad \dot{u} = -3H(u+p) < 0 \tag{6.9}$$

Because of $0 \leq u \leq 1$ (cf. figure 10), the inequalities (6.8) and, thus, (6.7) are satisfied. The equal sign is correct only for the start time $t = 0$, with $u = 1$. For all times $0 < t < \infty$, the particle horizon $D$ is greater than the linear extension $a$ of the universe. Causality between all regions of the universe during the expansion phase is ensured (figure 9).

*6.3 The flatness problem*

Let us look at the first Friedmann equation in (5.27) with the new radiation laws (5.28) in the spatial curved universe ($k = \pm 1$). In the system of units of (5.26), it reads

$$\left(\frac{\dot{a}}{a}\right)^2 + \frac{k}{a^2} = u \quad , \quad \dot{a} = \sqrt{u \cdot a^2 - k} \quad . \tag{6.10}$$

Let the development of all three cosmic versions $k = (0, \pm 1)$ at $t = 0$ begin with the same initial state, i.e. with the condition $a(0) = 1$. For $t \to 0$ ($\theta \to \infty$), then we have a convergence of $u \to 1$ and therefor $a \to 1$.

Now we examine the function $G = u \cdot a^2 > 0$. For $t = 0$, evidently $G(0) = 1$. Further, throughout the time range $0 \leq t < \infty$, the function $G(t)$ is falling strictly monotonously:

$$G = u\, a^2 = \dot{a}^2 + k \quad , \quad \dot{G} = 2\dot{a}\,\ddot{a} = \dot{a}[-(u+3p)a] = -H(u+3p)a^2 \tag{6.11}$$

With $\ddot{a}$ from (6.5) and the Hubble function $H > 0$ we obtain $\dot{G} < 0$ and, hence, the statement

$$0 \leq u \cdot a^2 \leq 1 \quad , \quad 0 \leq t < \infty \quad . \tag{6.12}$$

In the case of $k = +1$, the cosmological equation of motion reads

$$\dot{a} = \sqrt{u \cdot a^2 - 1} \quad . \tag{6.13}$$

Then, however, the Friedmann equation (6.13) cannot be satisfied. As also the initial velocity $\dot{a}(0) = 0$, a universe with curvature parameter $k = +1$ cannot develop with the new photon equations of state.

In the case of $k = -1$, the Friedmann equation (6.10) adopts the form

$$\dot{a} = \sqrt{u\,a^2 + 1} \qquad (6.14)$$

With the initial condition $a(t = 0) = 1$ and $u(t = 0) = 1$, the initial velocity for $t \to 0$ ($\theta \to \infty$) has the value $\dot{a}(t = 0) = \sqrt{2}$. Because of (6.11) the statement (6.12) is independent of the special value of $k$. Then the expansion rate $\dot{a}(t)$ with (6.14) keeps within the limits

$$1 \le \dot{a}(t) \le \sqrt{2} \quad , \quad 0 \le t < \infty \quad , \qquad (6.15)$$

and is, thus, permanently $> c$. Moreover, at the start of cosmic expansion is $(\dot{a}/a) > 1$, which contravenes the causality condition (6.7) or (6.8). Just at the beginning of cosmic evolution around $t \simeq 0$, however, the particle horizon $D$ should be greater than the linear dimension $a$ of the universe. For these physical reasons, we also want to reject an open universe with curvature parameter $k = -1$.

To sum up, we can conclude as follows: Should the new radiation law (5.11) or (5.15) be true, and should the early universe start to expand with a medium of the structure (2.2), this universe is flat from the beginning, with the curvature parameter $k = 0$ and the Friedmann equation

$$\dot{a}^2 = u\,a^2 \quad . \qquad (6.16)$$

Here, with (6.12), the expansion rate $\dot{a}$ is within the limits $0 \le \dot{a} \le 1$, i.e. it starts with $c$ and ends with $0$,

$$0 \le \dot{a} \le c \quad , \quad 0 \le t < \infty \quad . \qquad (6.17)$$

In this way, the flatness problem is solved by the structure of the equation of state: The equations of motion of the early universe (2.8) to (2.10), together with the photon equations of state (5.7) to (5.10) describe a flat continuously expanding universe with $k = 0$ according to equation (6.16). The three graphs in figure 10 for universes of type $k = 0, \pm 1$ illustrate this statement once more.

## 7 Concluding remarks

In his earlier report [2], the author derived a corrected Planck's radiation law for the energy density $u(T)$ from the hypothesis of the maximum quantum leap $\Delta E \le E_* = M_* c^2$ ($M_*$ Planck mass). The most important property of this radiation law for use in cosmology is the limit $\lim\limits_{T \to \infty} u(T) = u_\infty = (2/3) u_*$ ($u_*$ Planck energy density). Since, for $k_B T \ll E_*$, the new energy density formula $u(T)$ changes into the Stefan-Boltzmann law of classical radiation theory, deviations from the classical cosmological standard model of a radiation universe are expected only for the case of extremely high temperatures $k_B T \gg E_*$. Indeed we obtain, for this case in cosmic evolution, a flat universe expanding from a Planck region, which after some 100 Planck times smoothly joins the course of expansion of the FLRW standard model (figure 6). These primordial starting data of the universe at $t = 0$, obtained in the classical way, are as follows:

Volume $\quad V_0 = (4\pi/3)\, a_0^3 = 27.83\, L_*^3$, $\quad$ Velocity $\quad \dot{a}_0 = c$,

Energy density $\quad u_\infty = (2/3)\, u_*$, $\quad$ Pressure $\quad p \to \infty$,

Temperature $\quad T \to \infty$, $\quad$ Entropy $\quad S_\infty = 1.11\, k_B$.

After the primordial start of expansion, the particle horizon of the flat universe thus originating is always greater than the linear extension of the universe. The flatness problem itself is solved trivially, because with the new equation of state of the photon gas a closed or open universe with $k = \pm 1$ obviously comes out of question.

We suppose: Every universal limitation of the size of energetic quantum leaps in the theoretical foundations of quantum theory yields in the start phase of cosmic evolution an avoidance of singular initial states for the scale factor $a(t = 0) \neq 0$, $\dot{a}(t = 0) \neq 0$.

We note, in addition, that the new radiation formula (5.11) at first varies with scale. Invariance begins, when the cosmic temperature scale reaches $kT \ll E_*$, i.e. when the Stefan-Boltzmann law applies with sufficient accuracy.

Is there an entropy problem? According to (5.43), the cosmic expansion triggered off by the photon gas and dark energy is, in the case of equilibrium, an isentropic process with $S = S_\infty = const$. We can compute the entropy constant $S_\infty$ by further executing (5.43) and consider it in passage to the limit $T \to \infty$ ($t \to 0$). We start with

$$S = V \cdot s = V \cdot \frac{dp}{dT} = \frac{k_B}{E_*} V\, p' \quad , \quad p' = \frac{dp}{d\theta} \quad , \quad \theta = \frac{k_B T}{E_*} \quad . \tag{7.1}$$

As the $a(t)$ solution (5.42) of the Friedmann equation is given in terms of $[a_0, t_0, u_\infty]$, the formulas (7.1) needs to be converted into this system of units. Considering

$$V_* = 2\pi^2 L_*^3 \quad , \quad a_0 = \sqrt{\frac{9\pi}{8}} L_* \quad , \quad u_* = \frac{E_*}{V_*} = \frac{3}{2} u_\infty \tag{7.2}$$

we get, in continuation of (7.1),

$$S = k_B \frac{V}{V_*} \frac{p'}{u_*} = \frac{2}{3} k_B \frac{V}{V_*} \frac{p'}{u_\infty} = \frac{2}{3} k_B \left(\frac{2}{3\pi}\right)\left(\frac{a}{L_*}\right)^3 \frac{p'}{u_\infty} = \frac{k_B}{2} \sqrt{\frac{9\pi}{8}} \left(\frac{a}{a_0}\right)^3 \frac{p'}{u_\infty} \quad . \tag{7.3}$$

Now we can pass to the units of (5.26) and, with (5.22), substitute the solution (5.42) into the last formula:

$$S = \frac{k_B}{2}\sqrt{\frac{9\pi}{8}}\left(a^3 p'\right) = \frac{k_B}{2}\sqrt{\frac{9\pi}{8}}(3C_\infty) = \frac{9}{4}\sqrt{\frac{\pi}{2}}\left(\frac{\pi^2}{6} - \frac{5}{4}\right)k_B = 1.114\,k_B = S_\infty = const \quad . \quad (7.4)$$

It follows that *the pure radiation universe* in its early radiation phase at $t \to 0$ or $T \to \infty$ starts with the least possible entropy $S \simeq k_B$ and at first begins to expand adiabatically. This result confirms our conjecture that a modified electrodynamic radiation law at extremely high temperatures may lead to an avoidance of a cosmological big bang.

Current cosmological data, including estimations for start and end of a possible inflationary leap in the early phase of cosmic expansion connected with an entropy increasing can be found in [31].

Finally, one should be aware that the Friedmann equation (6.10) is time-symmetric, so that in the interval $-\infty < t \leq 0$ there could also exist the reverse process of a contraction of the universe down to the Planck size $a_0 = \sqrt{9\pi/8}\,L_*$.

Obviously, a better description of cosmological situations around the Planck region cannot be accomplished by the Friedmann equations derived from Einstein's classical gravitation theory. Here, a complete theory of quantum gravitation with applications to cosmology would be helpful.

**Appendix**

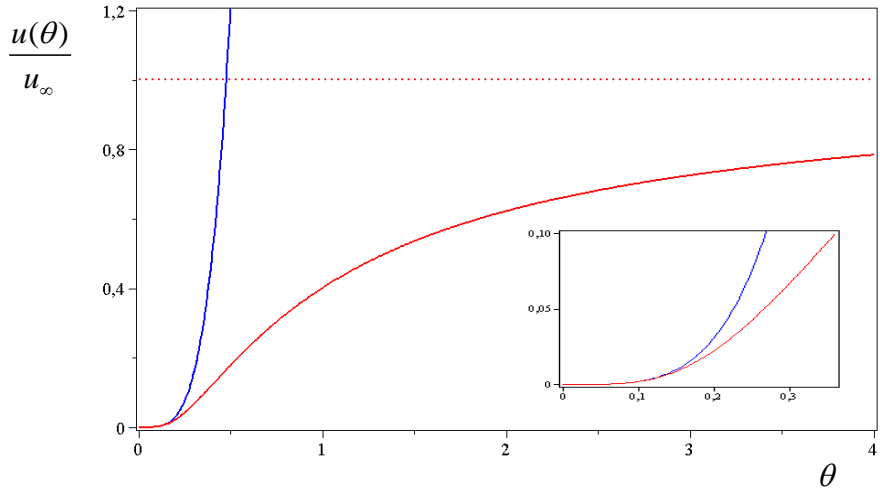

Fig. 1: The internal energy density of the photon gas $u(\theta)$ as a function of the temperature parameter $\theta = k_B T/E_*$. The blue curve follows the Planck course. In the temperature range up to about $k_B T \sim E_*/10$, the two curves are almost congruent, as shown by the inset. The new, red course of the function has, for extremely high temperatures $\theta = k_B T/E_* \to \infty$, a limit $u(\theta)/u_\infty \to 1$ (with $u_\infty = (2/3)u_*$).

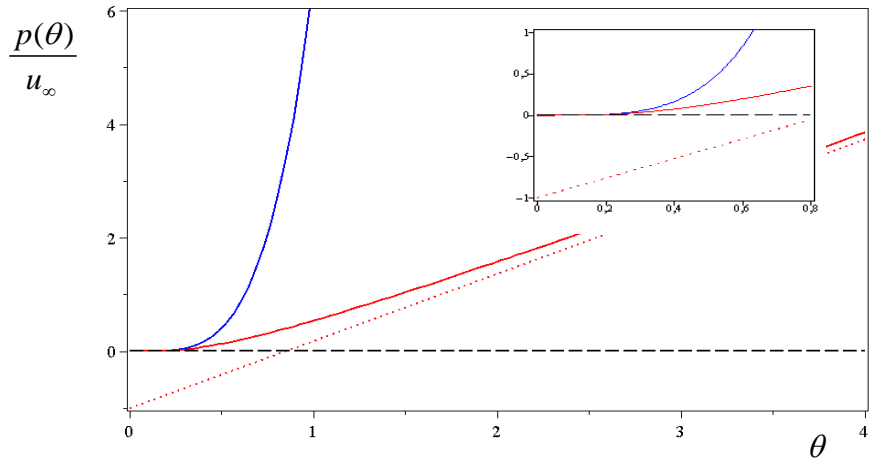

Fig. 2: The pressure of the photon gas $p(\theta)$ as a function of the temperature parameter $\theta = k_B T/E_*$. The blue curve follows the Planck course. In the temperature range up to about $k_B T \sim E_*/10$, the two curves are almost congruent, as shown by the inset. The new, red course of the function has, for extremely high temperatures $\theta = k_B T/E_* \to \infty$, an asymptote $p(\theta)/u_\infty = 3C_\infty \cdot \theta - 1$ (with $3C_\infty \approx 1.18$).

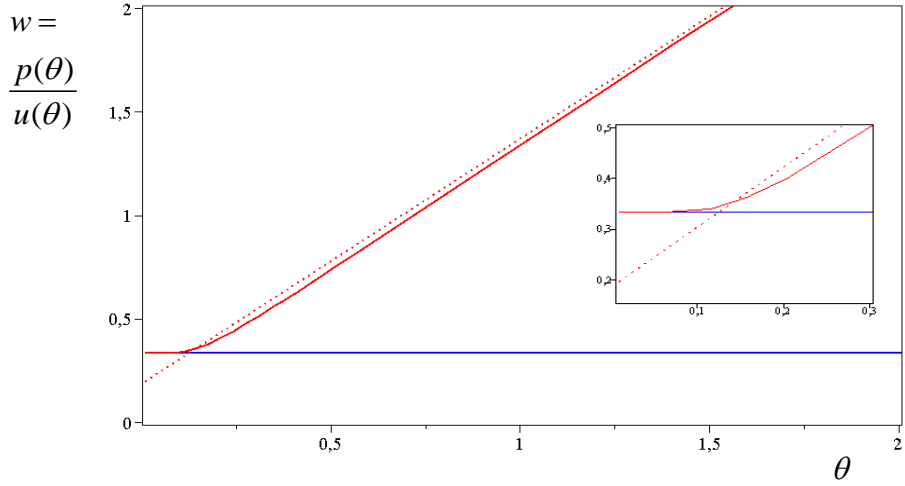

Fig. 3: The ratio $w(\theta) = p(\theta)/u(\theta)$ as a function of the temperature parameter $\theta = k_B T/E_*$. The blue curve follows the constant Planck course $w = p(\theta)/u(\theta) = 1/3$. In the temperature range up to about $k_B T \sim E_*/10$, the two curves are almost congruent, as shown by the inset. The new, red course of the function shows, for extremely high temperatures, a linear rise with the asymptote $w = p(\theta)/u(\theta) = 3C_\infty \cdot \theta + (3C_\infty - 1)$ (with $3C_\infty \approx 1.18$).

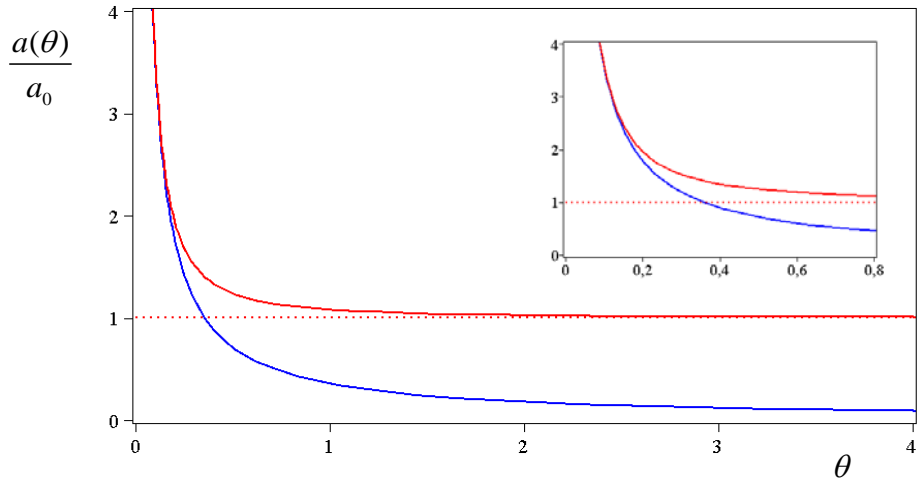

Fig. 4: The scale parameter $a(\theta)$ as a function of the temperature parameter $\theta = k_B T/E_*$. The blue curve follows the Planck course. In the temperature range up to about $k_B T \sim E_*/10$, the two curves are almost congruent, as shown by the inset. The new, red course of the function shows, for extremely high temperatures, a limit $a(\theta) \to a_0$ (with $a_0 = 1.88 L_*$).

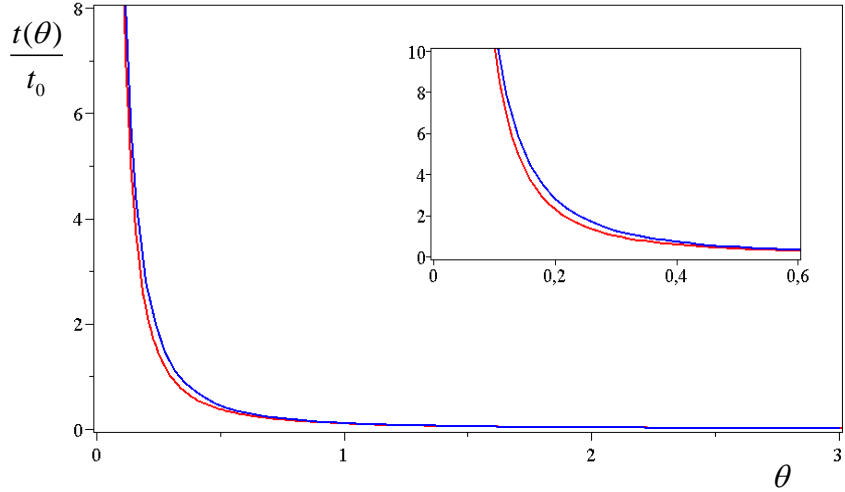

Fig. 5: The cosmic time $t(\theta)$ as a function of the temperature parameter $\theta = k_B T/E_*$. The blue curve follows the Planck course. Astonishingly, the two curves are almost congruent throughout the temperature range. As shown in the inset, there are slight deviations around $k_B T \sim E_*/5$. The time $t$ is plotted as a multiple of $t_0 = 1.88\, t_*$.

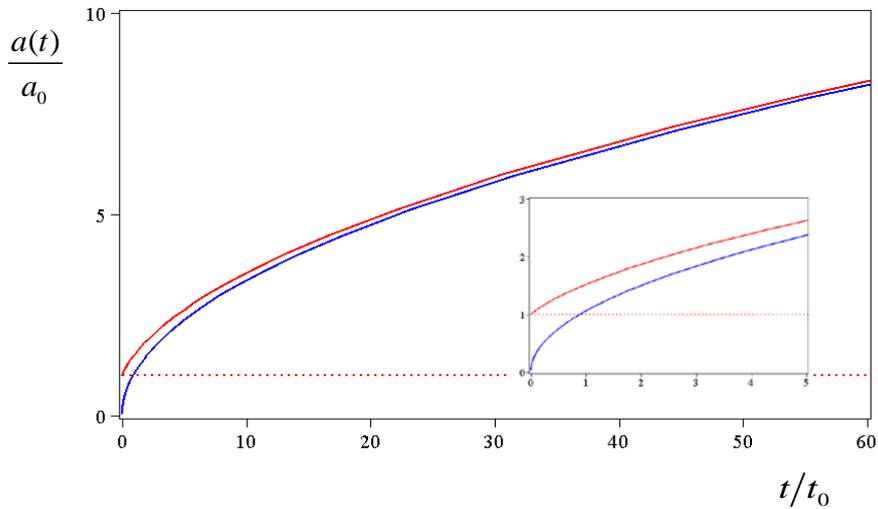

Fig. 6: The scale parameter $a(t)$ as a function of cosmic time $t$. The blue curve follows the Planck course. From a point in time around $t/t_0 \sim 50$, the two curves differ by only around one percent (cf. table 6.1). For the new, red function curve, the primordial start phase at time $t = 0$ begins at a finite value $a(t = 0) = a_0$ (with $a_0 = 1.88\, L_*$).

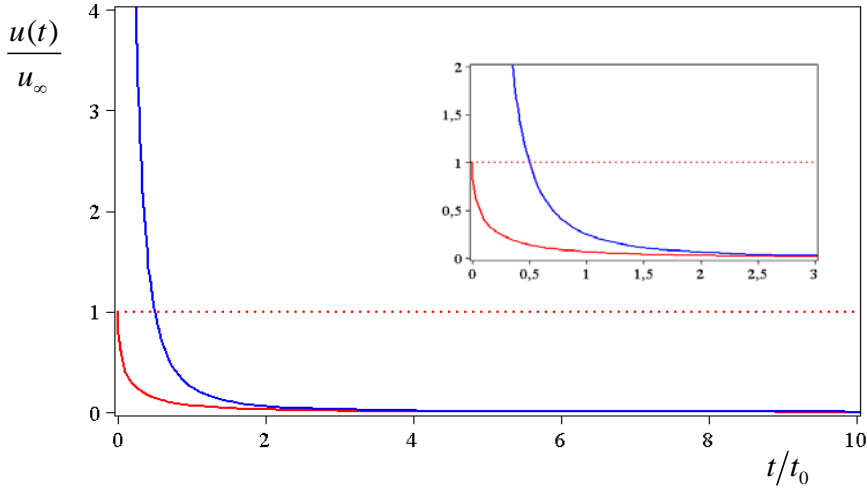

Fig. 7: The internal energy $u(t)$ as a function of cosmic time $t$. The blue curve follows the Planck course. From about $t > 3 t_0$, the two curves are almost congruent, as shown by the inset. For the new, red function curve, the primordial start phase at time $t = 0$ begins at a finite value $u(t = 0) = u_\infty$ (with $u_\infty = (2/3) u_*$ ).

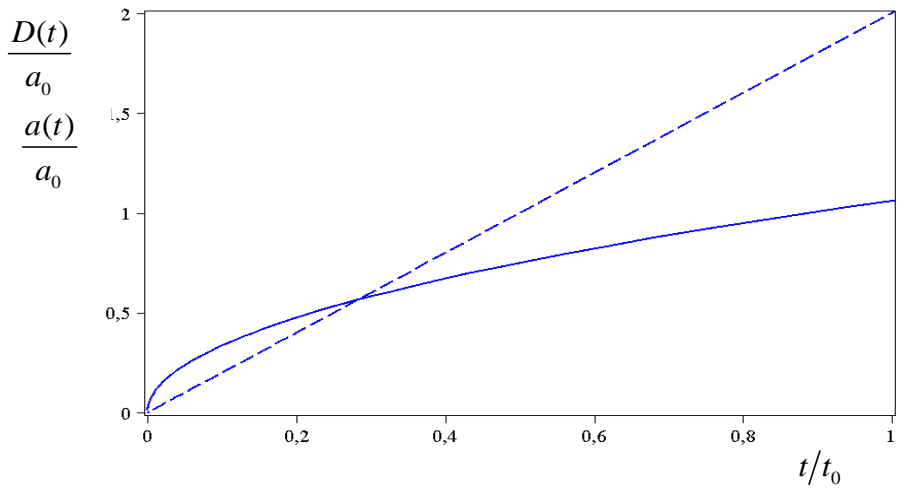

Fig. 8: The particle horizon $D(t)$ (dashed line) and the linear extension $a(t)$ of the universe (solid line) as a function of cosmic time $t$ in the case of Planck's radiation law. For $t \to 0$, there exists a point in time from which the particle horizon $D$ is smaller than the extension $a$ of the universe. From this time, obviously, causality throughout the universe, as a prerequisite for establishing homogeneity and isotropy of the cosmic microwave background, is no longer ensured.

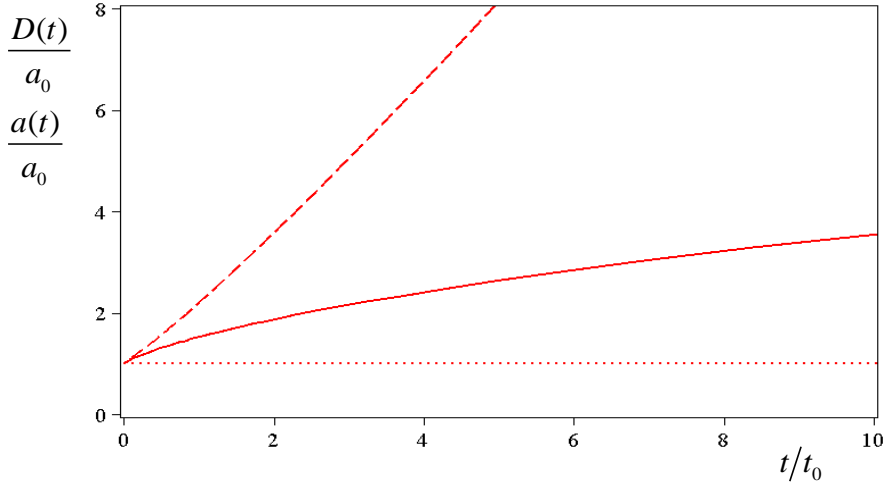

Fig. 9: The particle horizon $D(t)$ (dashed line) and the linear extension $a(t)$ of the universe (solid line) as a function of cosmic time $t$ in case of the new radiation law. Horizon $D$ and extension $a$ start their evolution in time at $t = 0$ with the start value $D(t=0) = a(t=0) = a_0$. For all times $t > 0$ remains invariably $D(t) > a(t)$. Causality is ensured during the expansion of the universe.

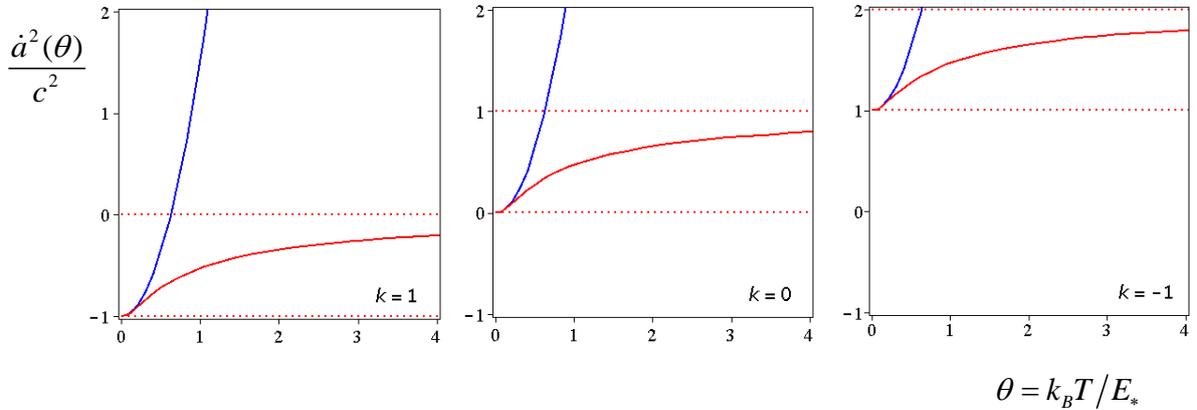

Fig. 10: The function $\dot{a}^2(\theta) = u\,a^2 - k$ versus the temperature parameter $\theta = k_B T/E_*$. The blue curves follow the Planck course. It is easy to see that, in the case of $k = +1$, the new, red curve is completely in the negative range $[-1, 0]$. With the new radiation law, therefore, a closed universe is ruled out from the start. In the Planckian case, the blue curve intersects the zero line and thus creates the possibility of expansion. In the case of $k = -1$, the two velocity curves are located above $\dot{a}^2 = c^2$. The only reasonable curve, obviously, is the red one in the case of $k = 0$, for which $0 \leq \dot{a}^2 < c^2$.

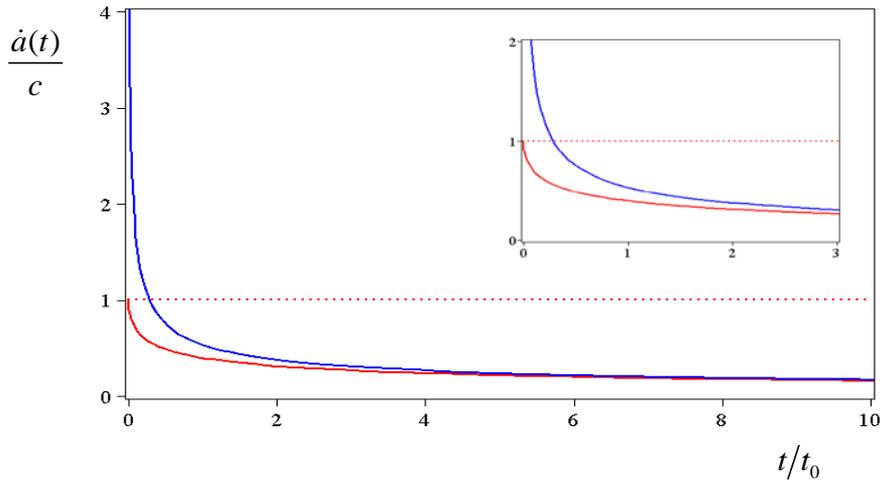

Fig. 11: The rate of expansion $\dot{a}(t)$ as a function of cosmic time $t$. The blue curve follows the Planck course. For the new, red curve, the primordial start begins with the finite initial velocity $\dot{a}(t=0) = c$ ($c$ velocity of light in vacuum). The inset shows the fast concourse of the two curves already for small expansion times $t > 3\,t_0$ (with $t_0 = 1.88\,t_*$).

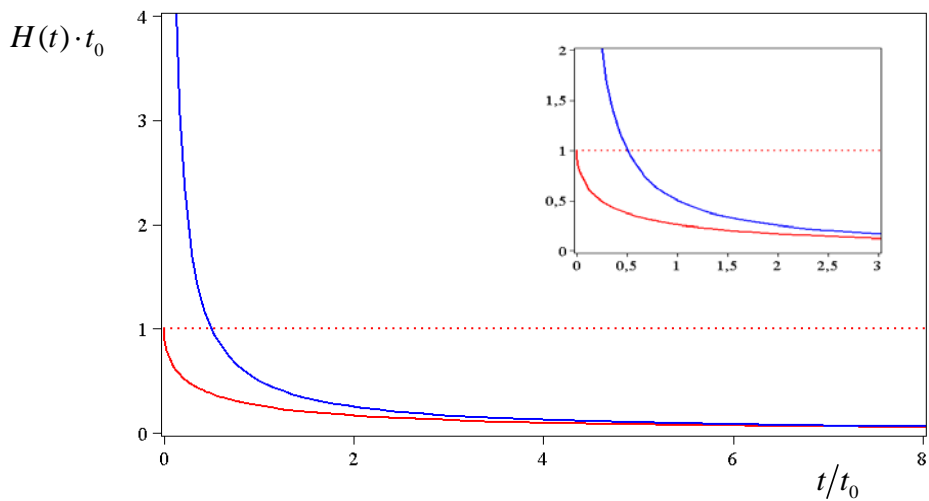

Fig. 12: The Hubble parameter $H(t)$ as a function of cosmic time $t$. The blue curve follows the Planck course. For the new, red curve, the primordial start begins with a finite Hubble value of $H(t=0) = 1/t_0$ (with $t_0 = 1.88\,t_*$). The inset shows the good conformity of the two curves already for expansion times $t > 3\,t_0$.